\documentclass[12pt]{iopart} 
\usepackage{graphicx}
\usepackage[psamsfonts]{amssymb}
\usepackage{calrsfs}
\usepackage{cite}
\renewcommand{\d}{{\rm d}}
\newcommand\Oscr{{\cal O}}
\newcommand{\eqref}[1]{\textup{({\ref{#1}})}}
\newcommand\twoR{\,{}^{\scriptscriptstyle{(2)}}\negthinspace R}

\newcommand\twog{\,{}^{\scriptscriptstyle{(2)}}\negthinspace g}
\def\twoR{\,{}^{\scriptscriptstyle{(2)}}\negthinspace R}

\def\dR{\,{}^{\scriptscriptstyle{(d)}}\negthinspace R}

\newcommand{\Hbb}{{\mathbb H}} 

\newcommand{\Ebb}{{\mathbb E}}
\newcommand{\xhor}{x_{\!\scriptscriptstyle\cal{H}}}
\newcommand{\rhor}{r_{\!\!\scriptscriptstyle\cal{H}}}

\begin{document}
\title[Poincar\'e ball embeddings]{Poincar\'e ball embeddings of the 
 optical geometry}
\author{M A Abramowicz\dag, I Bengtsson\S, V Karas\ddag\ and K Rosquist\S}
\address{\dag~Institute for Theoretical Physics, G\"oteborg~University and 
 Chalmers University of~Technology, S-412\,96~G\"oteborg,~Sweden}
\address{\ddag~Astronomical Institute, Charles~University,
 CZ-180\,00~Prague, Czech~Republic}
\address{\S~Department of Physics, Stockholm~University,
 Box~6730, S-113\,85~Stockholm, Sweden}
\ead{marek@fy.chalmers.se; ingemar@physto.se; vladimir.karas@mff.cuni.cz; 
 kr@physto.se}
\begin{abstract}
\noindent{}It is shown that optical geometry of the Reissner-Nordstr\"om 
exterior metric can be embedded in a hyperbolic space all the 
way down to its outer horizon. The adopted embedding procedure
removes a breakdown of flat-space embeddings which occurs outside the
horizon, at and below the Buchdahl-Bondi limit ($R/M=9/4$ in the
Schwarzschild case). In particular, the horizon can be captured in the
optical geometry embedding diagram. Moreover, by using the compact
Poincar\'e ball representation of the hyperbolic space, the embedding
diagram can cover the whole extent of radius from spatial infinity down
to the horizon. Attention is drawn to advantages of such embeddings
in an appropriately curved space: this approach gives compact embeddings
and it distinguishes clearly the case of an extremal black hole from a
non-extremal one in terms of topology of the embedded horizon.
\end{abstract}
\submitto{\CQG}
\maketitle


\section{Introduction}
This paper introduces a new, simple and convenient way to discuss physics of
spacetime near the black hole horizon in terms of the optical geometry
embedded in a Poincar\'e ball. We do not report a progress in developing new 
physical ideas. Instead, the goal of the present paper is more modest and
concentrated on a useful though quite inconventional embedding procedure. 

Embedding of curved spaces and spacetimes in a Euclidean space with a
higher number of dimensions is a well-known technique, so often used in 
general relativity that there is no need to recall its usefulness here
\cite{mtw:gravitation}.
The optical geometry is also useful, but perhaps less known. In a static
spacetime, it is defined by a particular conformal
rescaling of the three-dimensional geometry of space (i.e.\ a hypersurface
orthogonal to timelike Killing trajectories) in such a way that 
light trajectories {\it in space\/} are geodesic lines  
\cite{abramowicz_etal:1988}. Therefore, the
geometry of space can be directly established in terms of measurements based
solely on light tracing. This is a rather useful property of the optical
geometry that has already
been employed by numerous authors for remarkable simplifications of various
arguments and calculations, ranging from gravitational wave modes trapped in
super-compact stars \cite{sonego:2000} to origin of the Hawking radiation 
\cite{abramowicz:1997}.

Not only are geodesic lines optically straight, but they 
are also inertially, dynamically and electrically straight and their
direction agrees at each point with dynamical, inertial and electrical
experiments (e.g., a gyroscope that moves along such a straight line
does not precess). These properties follow from the rather remarkable fact
that all the relevant equations -- geodesic, Fermi-Walker, Maxwell,
Abraham-Lorentz-Dirac, Klein-Gordon -- when written in the 3+1 form of
the optical geometry are found to be identical with the corresponding
equations in Minkowski spacetime with a scalar field $\Phi$ (the
gravitational potential). For these reasons the geometry of optical
space offers simple explanations of several physical effects in strongly
curved spacetimes that otherwise could appear unclear or even confusing.
See ref.~\cite{abramowicz:1993} for a general exposition of the properties
of optical geometry, and ref.~\cite{sonego:1998} for a thorough discussion
and derivations.

The optical geometry of a given spacetime cannot, in general, be 
embedded in a Euclidean space all the way
to the horizon because of two separate reasons: (i)~the horizon is,
obviously, at an infinite distance in the optical geometry based on light
tracing; (ii)~optical geometry has a negative curvature near the horizon.
In this paper we show, by embedding the optical geometry in a Poincar\'e
ball, how one can avoid both of these difficulties.
The Poincar\'e ball embeddings show clearly the topology of the horizon of
the black hole. This feature should be helpful in discussing the role of the
horizon topology in the context of the Hawking radiation.
Our embeddings illustrate in a striking way the passage from
non-extremal to the extremal Reissner-Nordstr\"om hole. In
particular, the special nature of the extremal Reissner-Nordstr\"om
hole is clearly manifested in terms of a change in the topology of
the horizon. This feature is of interest in the context of
supersymmetric theories where the extremal black holes show up as
supersymmetric configurations \cite{horowitz}.

\section{An embedding procedure}
First, we briefly describe a way to embed a two-dimensional
hypersurface of a static and spherically symmetric spacetime in the
three-dimensional (curved) space of a \emph{Poincar\'e ball}
\cite{abhp:antidesitter,thurston:1998}. While a traditional approach to
visualization of spherically symmetric geometries employs a direct embedding
in the flat Euclidean space $\Ebb^3$, the present method does not 
suffer from several drawbacks that hamper
usual examples, such as the case of a Schwarzschild black hole
(\cite{mtw:gravitation}, chapt.\ 23). The region of spatial infinity is
mapped onto a compact interval in the embedding diagram, and at the same
time the whole range of radii down to the horizon is captured. 
We apply our method to the optical geometry of the Schwarzschild and 
Reissner-Nordstr\"om spacetime. We show
that this approach is particularly convenient in relation to the optical
geometry, and we argue that the advantages reach beyond the
visualization of the curved geometry.
  
The line element of a static, spherically symmetric spacetime can be
written in the form
\begin{equation}
 g = e^{2\Phi(r_{\!\star})}\Bigl[-\d t^2 + \d r_{\star}^2 + \tilde{r}^2(r_{\star})
              \left(\d \theta^2 + \sin^2\!\theta\, \d \phi^2\right) \Bigr]\,.
\end{equation}
The 3-space section $t={\it{}const}$ of the conformal metric (i.e.\ the
spatial part of the metric $\tilde{g}_{ik}=e^{-2\Phi}g_{ik}$ in the
square brackets) is called the optical space.  The variable $r_{\star}$ is
the Regge-Wheeler tortoise coordinate which plays the role of
geodesic radial distance in the optical geometry.

A convenient and often used representation of a curved geometry 
is by embedding it in flat Euclidean space. Take, for example, 
a 2-dimensional equatorial section $\theta=\pi/2$ of the optical
space
\begin{equation}
    g_{\scriptscriptstyle\Oscr}  = 
    \d r_{\star}^2 + \tilde{r}^2(r_{\star})\,\d\phi^2 \ .
    \label{eq:2d}
\end{equation}
We look for an isometric embedding of the 2-geometry \eqref{eq:2d} in the
Euclidean geometry
\begin{equation}
    g_{\mbox{\tiny{E}}}=\d\rho^2 + \rho^2\,\d\phi^2 + \d z^2 \ .
\end{equation}
To this aim, we define a height function $h(x){:=}z$ and a cylindrical
radius $q(x):=\rho$ where $x$ is an arbitrary, strictly monotonic
radial variable. The height and cylindrical radius functions together
give a parametric representation of the embedded surface.  By demanding
that the two angular coordinates coincide, the embedding becomes
uniquely determined by the embedding equations (see refs.\
\cite{mtw:gravitation} and \cite{spivak:1979} for a general discussion of
the embedding technique):
\begin{equation}
    \d r_{\star}^2 = \d q^2 + \d h^2 \ ,\qquad \tilde r = q \ .
\end{equation}
In these equations, $r_{\star}$ and $\tilde r$ should be considered as known
functions of $x$. Using the second equation to eliminate $q$, one obtains
\begin{equation}
   \frac{\d h}{\d r_{\star}}
            = \sqrt{1-\biggl(\frac{\d\tilde{r}}{\d r_{\star}}\biggr)^{\!2} } \ .
\end{equation}
This equation can be integrated if and only if the 
embedding condition,
\begin{equation}\label{eq:embed_cond}
    \biggl(\frac{\d\tilde{r}}{\d r_{\star}}\biggr)^{\!2} \leq 1 \ ,
\end{equation}
is fulfilled.  It should be noted that the above condition is necessary
and sufficient for the existence of an embedding which respects the
rotational symmetry of the optical geometry.  The existence of more
general embeddings (which we do not consider here) requires further
investigation. Both radial functions, $r_{\star}$ and $\tilde{r}$ appearing in
the optical metric, have clear geometrical meanings: Radial distances
are measured by $\d{r_{\star}}$ while distances along great circles are
measured by $\tilde{r}\,\d\phi$ at constant $r_{\star}$. For that reason one
refers to $r_{\star}$ as the \emph{optical radius}, and to $\tilde{r}$ as the
\emph{optical circumference} variable. Therefore, the content of the
embedding condition \eqref{eq:embed_cond} is such that the change in the
optical circumference cannot exceed the change in the optical radius.

The optical space of a spherically symmetric spacetime is completely
specified by the embedding function $F$ defined by $\tilde{r} = F(r_{\star})$.
In the case of the Schwarzschild geometry, the embedding function can be
written in the parametric form,
\begin{equation}
    \tilde{r} = \frac{r}{\sqrt{1-2M/r}} \ ,\qquad
	     r_{\star} = r+2M\ln\biggl( \frac{r}{2M}-1 \biggr) \ ,
\label{eq:rtilde}
\end{equation}
where $r$ is the usual Schwarzschild radius. The embedding condition
\eqref{eq:embed_cond} reads
\begin{equation}\label{eq:schwarz_cond}
   1 - \left(\frac{\d\tilde{r}}{\d r_{\star}}\right)^{\!2} =
    \frac{M\left(4r-9M\right)}{r\left(r-2M\right)} \geq 0 \ ,
\end{equation}
and it is satisfied for $r \geq 9M/4$ \cite{abramowicz_etal:1988}.
For a Reissner-Nordstr\"om black hole with electric charge $Q$, 
the embedding functions are given by the relations
\begin{equation}
  \tilde{r} = \frac{r^2}{\sqrt{r^2-2Mr+Q^2}}, \quad
  r_{\star} = r + w_+\ln\biggl(\frac{r}{r_+}-1\biggr)          
  -w_-\ln\biggl(\frac{r}{r_-}-1\biggr) 
  \label{eq:RN_rtilde}
\end{equation}
for $Q<M$, and
\begin{equation}
 \tilde{r} = \frac{r^2}{r-M} \ ,\quad
 r_{\star} = \frac{r(r-2M)}{r-M}
 + \ln\biggl(\frac{r}{M}-1\biggr)  
\end{equation}
for $Q=M$. Here, we denoted
\begin{equation}
w_\pm=\frac{M^2-Q^2/2}{\sqrt{M^2-Q^2}}\,{\pm}\,M,\quad
r_\pm=M\pm \sqrt{M^2-Q^2}.
\end{equation}
It follows from eq.~\eqref{eq:schwarz_cond} that the Schwarzschild
geometry can be embedded in flat Euclidean space only down to the
Buchdahl-Bondi limit, $R/M=9/4$.  This limit is also (coincidentally?) the 
lower limit for perfect fluid stellar equilibrium configurations in general
relativity. However, for a number of reasons, it is desirable to be able
to visualize curved geometries even beyond the Buchdahl-Bondi limit.
One way to achieve this goal is to perform the embedding in an
appropriately curved space. 

We start by writing the scalar curvature of
the optical geometry \eqref{eq:2d}, which is given by
\begin{equation}
    \twoR = -2F(r_{\star})^{-1} F''(r_{\star}) \ .
\end{equation}
For the Schwarzschild exterior solution, $\twoR=-2Mr^{-4}(2r-3M)$, and
$\d\!\twoR/\d r=12Mr^{-5}(r-2M)$. Hence the optical curvature becomes
more and more negative as we approach the horizon.  This leads us to
guess that we might be able to embed the optical space if the embedding
space has sufficiently large negative curvature. (This argument
should not be taken too literally.  For example, we know that an
$n$-sphere of arbitrarily large positive curvature can be embedded in
$\Ebb^{n+1}$.  It is ultimately the embedding condition
\eqref{eq:embed_cond} which determines whether the embedding is
possible.) The simplest and most natural choice is a space of constant
negative curvature, also known as hyperbolic space. In the next section
we review the basic properties of such spaces.

\section{The Poincar\'e ball as an embeddding space for the black hole 
 optical geometry}
\subsection{The Poincar\'e ball and the structure of a hyperbolic space}
\label{sec:pball}
Although the structure of hyperbolic space is quite well-known, it is
seldom a part of curricula of physics majors.  We are therefore
including a short discussion of the basics of hyperbolic geometry in
this section. Specifically, we discuss the symmetries and geodesic
structure of hyperbolic space, and the Poincar\'e ball in particular. As
mentioned above, a problem arises when embedding the optical geometry in
flat Euclidean space. In the case of a Schwarzschild black hole, the
embedding condition breaks down at $R=\case94M$.  One way out of the
difficulty is to use an embedding space which, like the Schwarzschild
optical geometry, possesses negative curvature, and a natural choice is
to use the Poincar\'e ball, a
special representation of $\Hbb^3$ (the 3-dimensional Riemannian space
of constant negative curvature, or, the hyperbolic 3-space for short).

By definition, hyperbolic space is a Riemannian space of constant
negative curvature. The metric of a 3-dimensional hyperbolic space can
be written in the form,
\begin{equation}\label{eq:pballmetric}
       g_{\mbox{\tiny\rm{H}}} = \frac{4\ell^2}{\left(1-\tilde\rho^2\right)^2}
                      \,\left(\d x^2 + \d y^2 + \d z^2  \right) \ ,
\end{equation}
where $\tilde\rho=\sqrt{x^2+y^2+z^2}$, and $\ell$ defines the radius of
curvature. The sectional curvature is given by $k=-1/\ell^2$, and this
relation remains valid also for the 2-dimensional hyperbolic space which
we will be using below. For constant curvature spaces, the sectional
curvature can be calculated from the Riemann curvature scalar $\dR$ by
the relation $\dR=kd(d-1)$ where $d$ is the dimension of the space. The
reader can recognize the metric \eqref{eq:pballmetric} as the spatial
part of the negative curvature Robertson-Walker geometry expressed in
isotropic coordinates. By identifying coordinates ($x,y,z$) with
Cartesian coordinates of $\Ebb^3$, the metric \eqref{eq:pballmetric} is
embedded (non-isometrically) in the sphere $\tilde \rho<1$.  This is the
Poincar\'e ball. Finally, we introduce a representation of hyperbolic
space known as the Poincar\'e half space. Its metric is
\begin{equation}\label{eq:poincareUH1}
 g_{\mbox{\tiny\rm{H}}} =
  \ell^2\zeta^{-2}\left(d\xi^2+d\eta^2+d\zeta^2\right) \ ,
\end{equation}
where $(\xi,\eta,\zeta)$ are Cartesian embedding coordinates in the upper
half space region $\zeta>0$ of $\Ebb^3$.

\subsection{Two representations of a hyperbolic space}
We recall the basic geometrical features characterizing a hyperbolic
space and its representations as the upper half space and the ball.
Although we need the 3-dimensional version of hyperbolic space for the
embeddings, it suffices to consider the Killing orbits and geodesics in
a 2-dimensional plane.  Once the 2-dimensional picture is clear, the
3-dimensional picture can easily be obtained by rotation.

\begin{figure}[!tb]
\begin{center}
\includegraphics[width=0.5\textwidth]{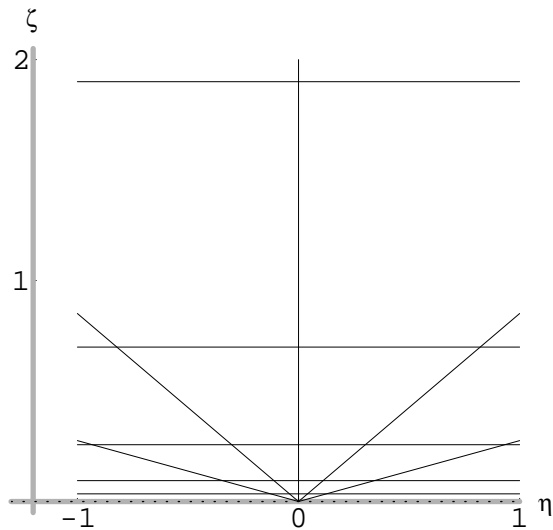}
\end{center}
\caption{Translational Killing orbits in Poincar\'e upper half
  space. The dotted line, $\zeta=0$, marks the boundary of the
  hyperbolic space.  The boundary is infinitely distant to any point in
  the interior ($\zeta>0$) of the space.  The solid horizontal lines are
  equidistant orbits of $k_1$.  The ray orbits correspond to $k_2$, and
  they too are equidistant.  In both cases the distance between the
  orbits is unity if $\ell=1$.  These orbits are also shown below in
  fig.~\ref{fig:disk-killing}.}
   \label{fig:uh-killing}
\end{figure}

We start with the upper half space which is mathematically simpler.  
A 2-dimensional restriction can be defined by setting $\xi=0$ in the 
metric \eqref{eq:poincareUH1},
\begin{equation}
 \twog_{\mbox{\tiny\rm{H}}} = 
 \ell^2\zeta^{-2}\left(\d\eta^2 + \d\zeta^2\right)\ .
\end{equation}
This metric has three independent Killing fields corresponding to two
translations and one rotation.  Two independent translations are defined 
in terms of Killing fields without fixed points in the following way:
\begin{equation}
    k_1=\frac{\partial}{\partial\eta} \ ,\quad 
    k_2=\eta\,\frac{\partial}{\partial\eta}
         +\zeta\,\frac{\partial}{\partial\zeta} \ .
\end{equation}
The integral curves of these fields are displayed in
fig.~\ref{fig:uh-killing}.  We refer to the class of orbits
corresponding to $k_1$ as the \emph{horizontal} class in the upper half
space, while the class given by $k_2$ may be referred to as the
\emph{ray} class.

\begin{figure}[!tb]
\begin{center}
\includegraphics[width=0.5\textwidth]{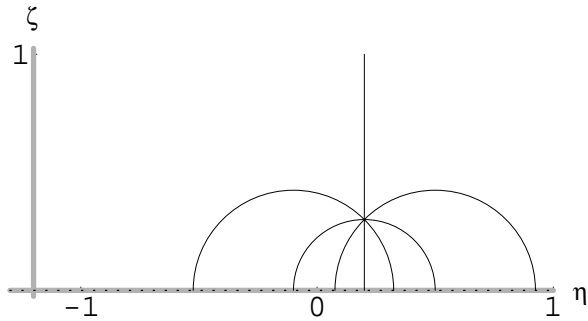}
\end{center}
\caption{Geodesics in Poincar\'e upper half space.  A sample of
  geodesics through a point at $\pi/4$ angular separation.  Note that the
  angles in $\Hbb^{2}$ coincide with the angles in $\Ebb^{2}$ due to the
  conformal equivalence between the two spaces.  }
    \label{fig:uh-geodesics}
\end{figure}

To understand fully the hyperbolic space geometry we must also know its
geodesics.  In contrast to the situation in flat space, the geodesics in
hyperbolic space do not coincide with Killing orbits. Returning to
the 2-dimensional upper half space picture, the geodesics are
semicircles with centers on the line $\zeta=0$ (see
fig.~\ref{fig:uh-geodesics}).  The vertical lines are also geodesics and
can be considered as a limiting case by moving the center of the circle
along $\zeta=0$  to infinity. It should be kept in mind that the special
appearance of the vertical geodesics is entirely due to the adopted
representation.  All geodesics are in fact equivalent due to complete
isotropy and homogeneity of the hyperbolic space.

Our next task is to transform the above statements about the upper half
space to the case of Poincar\'e disk. This can be achieved most
transparently by using M\"obius transformation to transform the upper
half plane with complex coordinate $w=\eta+\mbox{i}\zeta$ into a unit disk. The
required transformation is
\begin{equation}\label{eq:mobius}
    w \rightarrow w' := y + \mbox{i}z = \frac{\mbox{i}w+1}{w+\mbox{i}} \ .
\end{equation}
Relation between the disk coordinates, $(y,z)$, and the upper half space
coordinates, $(\eta,\zeta)$, can also be written in the real form
\begin{equation}
    y = \frac{2\eta}{\eta^2+\left(1+\zeta\right)^2} \ ,\qquad
    z = \frac{\eta^2+\zeta^2-1}{\eta^2+\left(1+\zeta\right)^2} \ .
\label{eq:ball_trans}
\end{equation}
This gives the 2-dimensional hyperbolic space in the Poincar\'e disk form
\begin{equation}
  \twog_{\mbox{\tiny\rm{H}}}=
   \frac{4\ell^2}{\left(1-\tilde\rho^2\right)^2}\,
   \left(\d y^2 + \d z^2\right)\ ,
\end{equation}
where $\tilde\rho^2 = y^2 + z^2$.

\begin{figure}[!tb]
\begin{center}
\hfill
\includegraphics[width=0.42\textwidth]{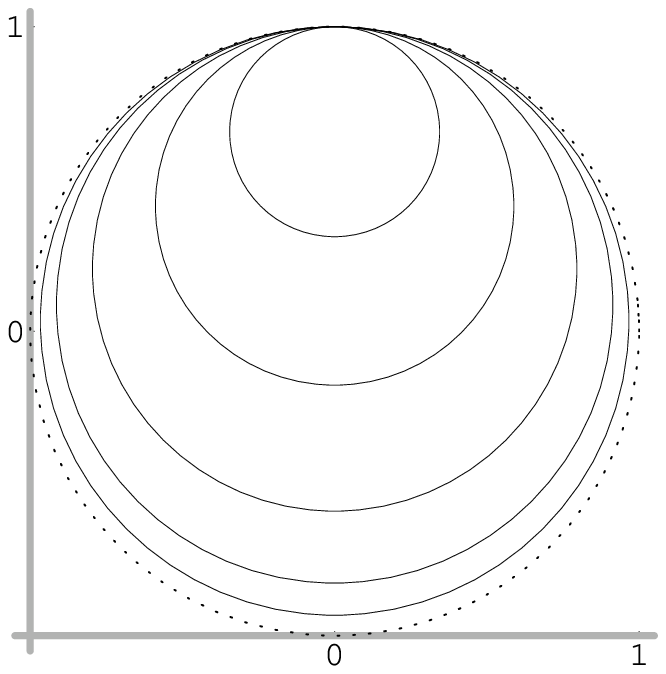}
\hfill
\includegraphics[width=0.42\textwidth]{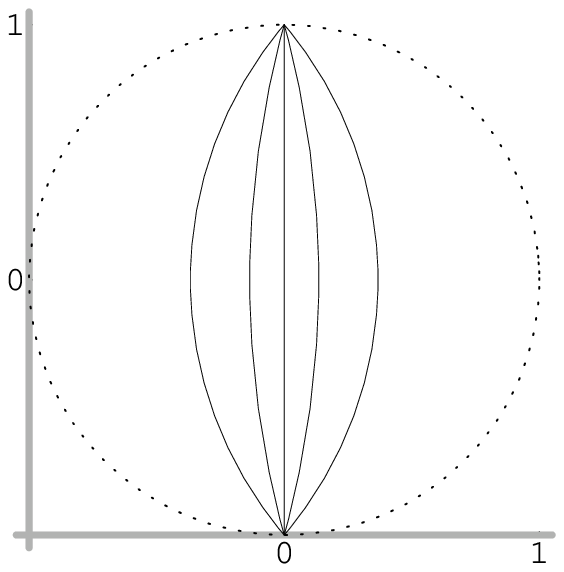}
\hfill
\end{center}
\caption{Two classes of Killing orbits are shown in the
  Poincar\'e disk, corresponding to the orbits plotted in
  fig.~\ref{fig:uh-killing}: The north pole class (left panel) is
  identical to the family of horizontal lines in
  fig.~\ref{fig:uh-killing}, while the meridional class (right panel)
  coincides with the family of ray orbits.  The dotted circle represents
  the boundary of $\Hbb^{2}$.}
   \label{fig:disk-killing}
\end{figure}

Let us start with the horizontal class of Killing orbits as given in
fig.~\ref{fig:uh-killing}. These orbits appear as circles touching the
north pole in the disk picture of fig.~\ref{fig:disk-killing} (left
panel). Clearly, the term {\it{}horizontal} does not apply to these
orbits in the context of the disk representation. Instead they will be
referred to as the \emph{north pole} class of Killing orbits.

\begin{figure}[!tb]
\begin{center}
\includegraphics[width=0.5\textwidth]{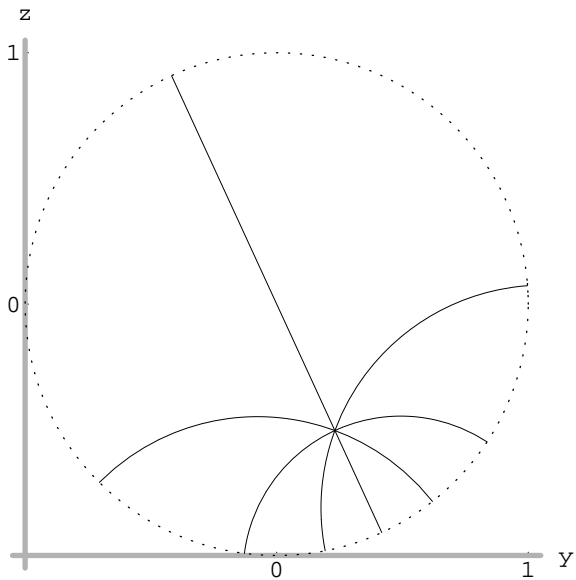}
\end{center}
\caption{Geodesics in the Poincar\'e disk.  A sample of geodesics
  at $\pi/4$ angular separation through the transformed reference point
  of fig.~\ref{fig:uh-geodesics}.  As explained in text, however, these
  are not the same geodesics as those in that figure.  }
   \label{fig:disk-geodesics}
\end{figure}

Next we consider, in more detail, the ray orbits from
fig.~\ref{fig:uh-killing}. They too are circular in the disk but with
the center of the circle lying outside the disk itself.  The entire
orbits are therefore represented by circular arcs inside the disk, and
their appearances do not resemble rays anymore. We will refer to them as
\emph{meridional} Killing orbits (see fig.~\ref{fig:disk-killing}, right
panel) because, in the 3-dimensional view, each such orbit gives rise to
a surface of revolution which defines a meridional Killing cylinder.

Finally, we take a look at the geodesics in the disk picture.  As in the
upper half space, the geodesics are circles which meet the boundary at
right angle, as illustrated in fig.~\ref{fig:disk-geodesics}.  The
straight lines going through the origin are geodesics that
represent the limiting case when radius of the circle is infinite.
The straight-line geodesics are in a one-to-one correspondence with
those upper half space geodesics which pass through the point $w=i$ via
eq.~\eqref{eq:mobius}. A sample of geodesics is shown in 
fig.~\ref{fig:disk-geodesics} (they have been constructed at $\pi/4$ angular 
separation). Note that the mapping \eqref{eq:mobius} transforms the geodesics 
of fig.~\ref{fig:uh-geodesics} to a different set of disk geodesics (also
at $\pi/4$ angular separation).

\section{Reissner-Nordstr\"om optical geometry embedded in a
 hyperbolic space}
\label{sec:hyperbolic_embeddings}
\subsection{The embedding equations}
\label{sec:embedding_equations}
When considering the embeddings in a hyperbolic space we need to find an
embedding condition which replaces eq.~\eqref{eq:embed_cond} valid for flat
space. For that purpose we
introduce cylindrical coordinates $(\tilde\chi,\phi,z)$ by
\begin{equation}
         x = \tilde\chi\cos\phi \ ,\qquad
         y = \tilde\chi\sin\phi \ .
\end{equation}
The Poincar\'e ball metric then takes the form
\begin{equation}\label{eq:pball_cyl}
    g_{\mbox{\tiny\rm{H}}} = \frac{4\ell^2}{\left(1-\tilde\rho^2\right)^2}
           \left(\d\tilde\chi^2 + \tilde\chi^2 \d\phi^2 + \d z^2\right) \, ,
\end{equation}
where $\tilde\rho^2 = \tilde\chi^2 + z^2$.  The equatorial part of the
optical geometry of a static spherically symmetric spacetime is given by
the 2-metric \eqref{eq:2d}.  Comparing with the ball metric in cylindrical
coordinates, eq.~\eqref{eq:pball_cyl}, we define the height function
$h:=z$ and the cylindrical radius function $q:=\chi = \sqrt{x^2 + y^2}$.
This gives the embedding relations
\begin{equation}\label{eq:embed_ball}
    \d r_{\star}^2 = \frac{4\ell^2}{\left(1-s^2\right)^2}
        \left(\d q^2+\d h^2\right) \ ,\qquad
    \tilde{r} = \frac{2\ell q}{1-s^2} \ ,
\end{equation}
where we have introduced $s^2 := q^2 + h^2$. Just as for flat space
embeddings there is an embedding condition which must be satisfied (this
hyperbolic embedding condition will be discussed in section
\ref{sec:hyper_embed}).  When $\tilde r$ is a known function of $r_{\star}$,
the equations \eqref{eq:embed_ball} can be solved providing the
embedding condition is satisfied.  We have solved these equations
numerically for some cases to be discussed below. Solving them
analytically is not straightforward. However, it
is possible to obtain further insight by first performing the embedding
in the upper half space and then transforming to the ball picture.  In
fact, it turns out that the embedding relations can be rather easily
solved analytically in that case.

We introduce cylindrical coordinates $(\chi,\phi,\zeta)$ for the upper half
space by 
\begin{equation}
 \xi  = \chi\cos\phi \ ,\qquad \eta = \chi\sin\phi \ 
\end{equation}
(we do not distinguish the angle coordinates in different
representations since we are only considering spaces with rotational
symmetry).
Writing the upper half space metric in cylindrical coordinates gives
\begin{equation}
    g_{\mbox{\tiny\rm{H}}}=\ell^2\zeta^{-2} 
      \left(\d\chi^2 + \chi^2\d\phi^2 + \d\zeta^2\right)\ .
\end{equation}
We define a height function $h(r)=\zeta$ and the cylindrical 
radius function $q(r)=\chi$ for the values of $\zeta$ and $\chi$ of 
the embedded surface.  The two functions are determined by the embedding 
equations
\begin{equation}\label{eq:h_embed_cond}
    \d r_{\star}^2 = \ell^2h^{-2}\left(\d q^2 + \d h^2\right) \ ,\quad
    \tilde{r} = \ell qh^{-1} \ .
\end{equation}
These relations have a structure simpler than the corresponding equations 
for the ball embedding \eqref{eq:embed_ball}.  To find the solution we 
first eliminate $\d{q}$ to obtain an equation which is quadratic 
in $(\ln h)'$:
\begin{equation}
  \left(\tilde{r}^2 + \ell^2\right)\left(\ln h\right)'{}^2
  + 2\tilde{r}\tilde{r}'\left(\ln h\right)'
  + \tilde{r}'{}^2 - r'_{\star}{}^2 = 0 \ ,
\end{equation}
where the prime stands for differentiation with respect to $r$.  
The solution is
\begin{equation}\label{eq:lnhprime}
 (\ln h)'=\frac{-\tilde{r}\tilde{r}'+ \sqrt{{\cal{}D}(r)}}{\tilde{r}^2+\ell^2} \ ,
\end{equation}
where
\begin{equation}
 {\cal{}D}(r):=\tilde{r}^2r'_{\star}{}^2+\ell^2\left(r'_{\star}{}^2-\tilde{r}'{}^2\right) \ ,
\end{equation}
and the positive root has been chosen to ensure $h'>0$.
Integrating eq.~\eqref{eq:lnhprime} we obtain
\begin{equation}\label{eq:hint}
    h = \frac{1}{\sqrt{\tilde{r}^2(r)+\ell^2}}
                      \exp\left\{ \int{\frac{\sqrt{{\cal{}D}(r)}\,\d r}
                      {\tilde{r}^2(r)+\ell^2}} \right\} \ .
\end{equation}
In the Reissner-Nordstr\"om case we have
\begin{equation}
  {\cal{}D}(r)=
  \frac{r^2\left\{ r^6 + \ell^2\left[4Mr^3-3\left(3M^2+Q^2\right)r^2+
   12MQ^2r-4Q^4\right] \right\}}{\left(r^2-2Mr+Q^2\right)^3} \ .
\end{equation}
Setting $x=r/M$, $\alpha=\ell/M$ and $\mu=Q/M$ 
we can now write the height function in the explicit form,
\begin{equation}\label{eq:h_expr1}
    h(x) = K\, h_\infty f(x)
             \exp\left\{ \int_{\xhor}^x g(u)\,\d u \right\} \ , 
\end{equation}
where
\begin{eqnarray}
  f(x) &= &\frac{1}{\sqrt{\tilde{r}^2(r)+\ell^2}}
        = \sqrt{\frac{x^2-2x+\mu^2}{x^4+\alpha^2(x^2-2x+\mu^2)}} \ , \nonumber \\
  g(x) &= &\frac{\sqrt{{\cal{}D}(r)}}{\tilde{r}^2(r)+\ell^2}
        = \frac{x\sqrt{x^6+\alpha^2[4x^3-3(3+\mu^2)x^2+12\mu^2x-4\mu^4]}}
                {[x^4+\alpha^2(x^2-2x+\mu^2)]\sqrt{x^2-2x+\mu^2}} \ ,
\end{eqnarray}
and $\xhor = 1+ \sqrt{1-\mu^2}$ is the value of $x$ at the horizon. The 
expression \eqref{eq:h_expr1} cannot be used as it stands to evaluate the 
height function. The reason is that the integral diverges in the limit
$x\rightarrow\infty$ as will be discussed below. Moreover, for $Q=M$, the
integral also diverges in the limit $x\rightarrow\xhor$. We
now proceed to discuss the case $Q<M$, while the extremal case $Q=M$ can be
dealt with along similar lines.

The value of the constant $K$ is assumed to be chosen such that $h_\infty$ is
the asymptotic value of the height function in the limit $x\rightarrow
\infty$.  To see how this works, we must understand the behaviour of $h$ in
that limit. We first note that $f(x)$ and $g(x)$ have the asymptotic forms
\begin{equation}
    f(x) = \frac1x + \Oscr\left(\frac1{x^2}\right) \ ,\qquad
    g(x) = \frac1x + \frac1{x^2} + \Oscr\left(\frac1{x^3}\right) \ .
\end{equation}
Defining
\begin{equation}\label{eq:S_asymp}
    {\cal{}S}(x) := xg(x)-1 = \frac1x + \Oscr\left(\frac1{x^2}\right) \ ,
\end{equation}
we can then express $h$ in the form
\begin{equation}\label{eq:h_expr2}
    h(x) = K\,h_\infty\,\xhor^{-1} x f(x) \exp\left\{\int_{\xhor}^x
                \frac{{\cal{}S}(u)\d u}{u} \right\} \ .
\end{equation}
Now, using the fact that $xf(x) \rightarrow 1$ as $x \rightarrow \infty$, 
we find that
\begin{equation}\label{eq:K_expr}
    K = \xhor \exp\left\{ -\int_{\xhor}^\infty
              \frac{{\cal{}S}(u)\d u}{u} \right\} \ .
\end{equation}
It follows from the asymptotic form of ${\cal{}S}(x)$ that the integral in
\eqref{eq:K_expr} is convergent and hence $K$ is a well-defined positive
constant.  The embedded surface is asymptotic to the plane $h=h_\infty$.
The value of $h_\infty$ is arbitrary and reflects only a choice of scale
which affects both $h$ and $q$ in the same way as seen from the
embedding equations \eqref{eq:h_embed_cond}.  From \eqref{eq:RN_rtilde}
and \eqref{eq:h_embed_cond} we also obtain the cylindrical radius
function:
\begin{equation}\label{eq:chix}
    q(x) = \frac{x^2 h(x)}{\alpha\sqrt{x^2-2x+\mu^2}} \ .
\end{equation}
The relations \eqref{eq:h_expr1} and \eqref{eq:chix} together give the
embedding of Reissner-Nordstr\"om optical geometry in the upper half
space. The ball embedding functions can be readily computed from the upper
half space embedding using the coordinate transformation
\eqref{eq:ball_trans}:
\begin{equation}
 \tilde q = \frac{2q}{q^2+(1+h)^2} \ ,\quad
 \tilde h = \frac{q^2+h^2-1}{q^2+(1+h)^2} \ .
\end{equation}                  

\subsection{The hyperbolic space embedding condition}
\label{sec:hyper_embed}
A necessary condition for embedding is ${\cal{}D}(r)\geq0$. 
This criterion can be expressed in the dimensionless form
\begin{equation}\label{eq:Bdef}
    {\cal{}B}(r) := \frac{M^2}{\tilde{r}^2}\left[
         \left( \frac{\d\tilde{r}}{\d r_{\star}} \right)^{\!2} - 1 \right]
          \leq \frac{M^2}{\ell^2}\ .
\end{equation}
In the limit of flat space embeddings, $\ell \rightarrow \infty$, this
condition reduces to \eqref{eq:embed_cond} as it should.  It follows that
the possibility of performing an embedding is governed by the quantity $\sup
{\cal{}B}(r)$ where the supremum is taken over some desired radial range.  The
reasonable range in this context is $r> \rhor$ where $\rhor$ is the value of
$r$ at the horizon.  If $\sup {\cal{}B}(r) \leq0$, then $\Ebb^3$ can be used as
embedding space.  If, on the other hand, $0< \sup {\cal{}B}(r) < +\infty$ and 
if the radius of curvature satisfies the global embedding condition
\begin{equation}\label{eq:embed_global}
    \ell \leq M [\,\sup {\cal{}B}(r)]^{-1/2} ,
\end{equation}
then hyperbolic spaces can be used.
For the Reissner-Nordstr\"om solution we have
\begin{displaymath}
    {\cal{}B}(r) = M^2 r^{-6} \left[-4Mr^3 + 3\left(3M^2+Q^2\right)r^2
          - 12MQ^2r + 4Q^4 \right] \ .
\end{displaymath}
If we restrict attention to the case $Q\leq M$, then this function attains its
supremum at the (outer) horizon $\rhor = r_+ = M+\sqrt{M^2-Q^2}$ with the
value
\begin{equation}
   {\cal{}B}(\rhor) = \sup_{r>\rhor}{\cal{}B}(r)
            = \frac{1-\mu^2}{\Bigl(1+\sqrt{1-\mu^2}\Bigr)^4} \ ,
  \label{eq:lm}
\end{equation}
where $\mu:=Q/M$.  It follows that the global embedding condition
\eqref{eq:embed_global} takes the form
\begin{equation}\label{eq:RN_embed_cond}
    \ell/M \leq 2 + \frac{2-\mu^2}{\sqrt{1-\mu^2}} \ .
\end{equation}
Thus for the Schwarzschild metric, the global embedding condition is $\ell/M
\leq 4$.  The right hand side of \eqref{eq:RN_embed_cond} increases
monotonically with $\mu$ and obviously tends to infinity in the limit $\mu
\rightarrow 1$.  This implies that in the extreme Reissner-Nordstr\"om case, 
the optical geometry can in fact also be embedded in flat space.

\begin{figure}[!tb]
\begin{center}
\includegraphics[width=0.8\textwidth]{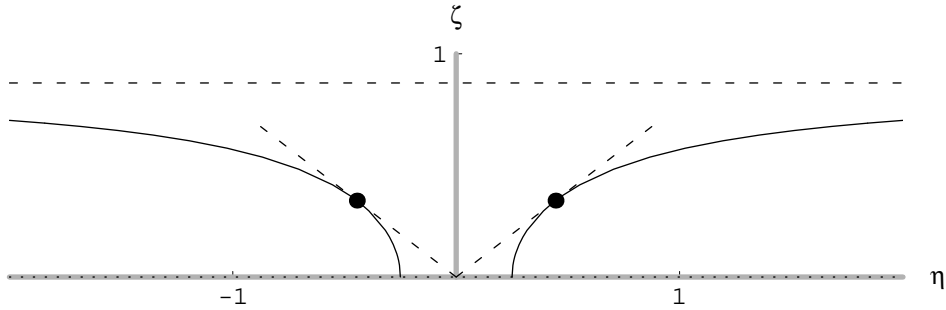}
\end{center}
\caption{A vertical section of the embedded Schwarzschild
  geometry in the upper half space.  As $r \rightarrow \infty$ the
  optical geometry is asymptotic to the Killing orbit represented by the
  dashed horizontal line. The neck at $r=3M$ is indicated by the black
  dots where the optical geometry touches the dashed Killing ray orbits.}
   \label{fig:uh-schwarzschild}
\end{figure}

\begin{figure}[!tb]
\begin{center}
\includegraphics[width=0.5\textwidth]{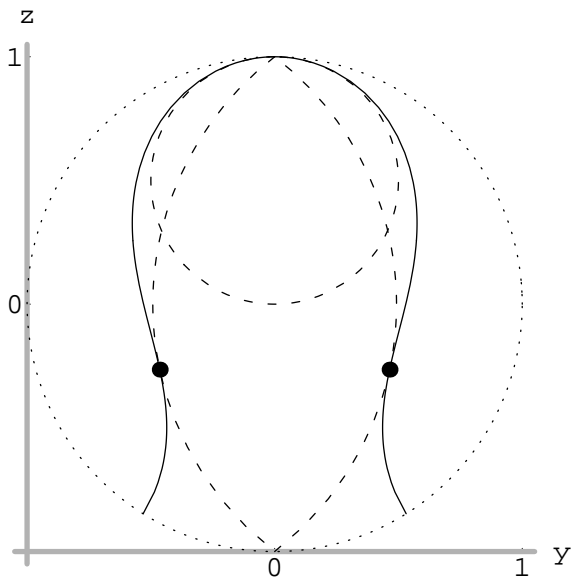}
\end{center}
\caption{A vertical section of the embedded Schwarzschild
  geometry in the Poincar\'e ball. The dashed curves represent Killing
  orbits. The black dots mark the position of the neck at $r=3M$. }
   \label{fig:disk-schwarzschild}
\end{figure}

\subsection{Physical properties of the embedding}
\label{properties}

\begin{figure}[!tb]
\begin{center}
\hfill
\includegraphics[width=0.4\textwidth]{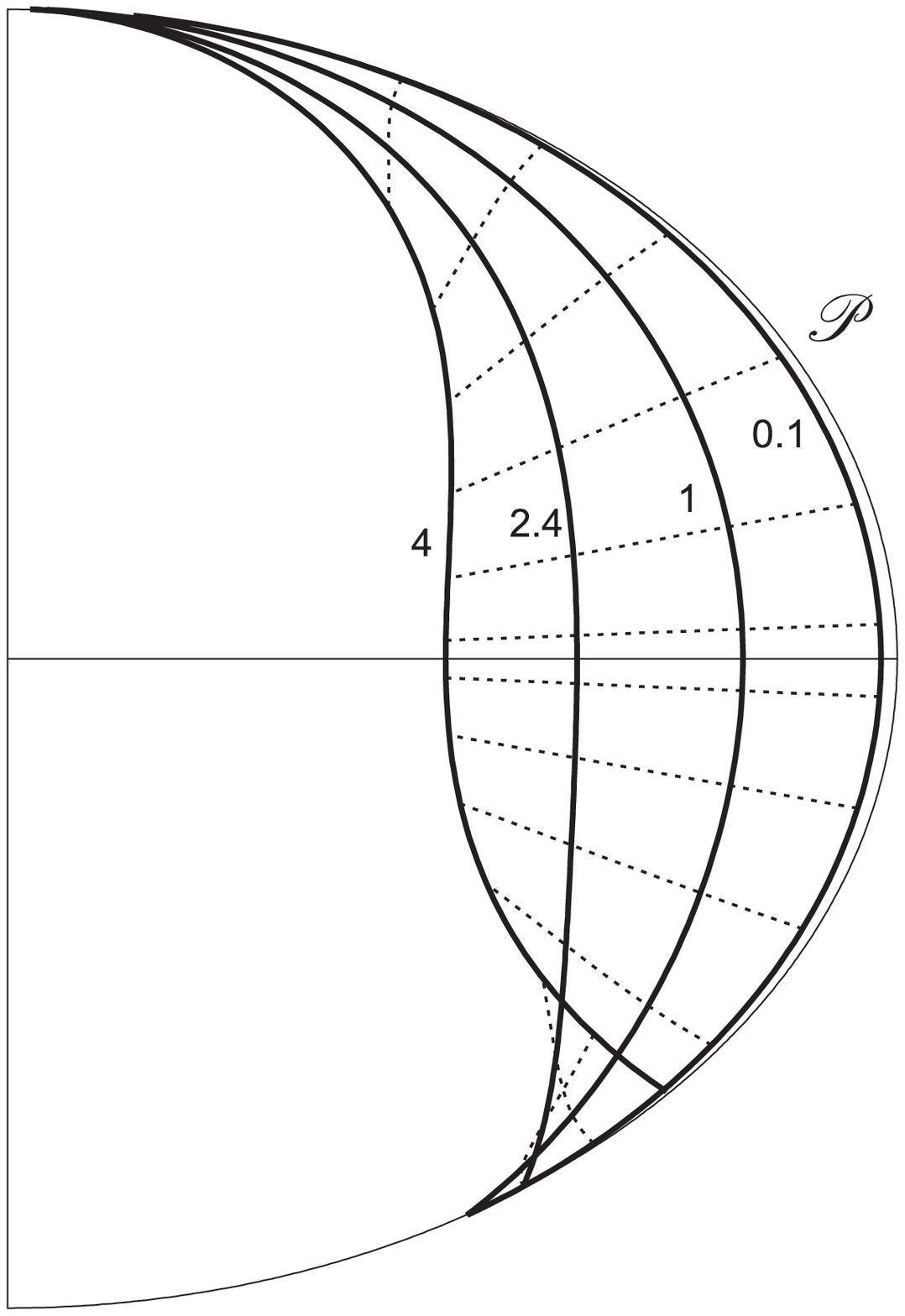}
\hfill
\includegraphics[width=0.4\textwidth]{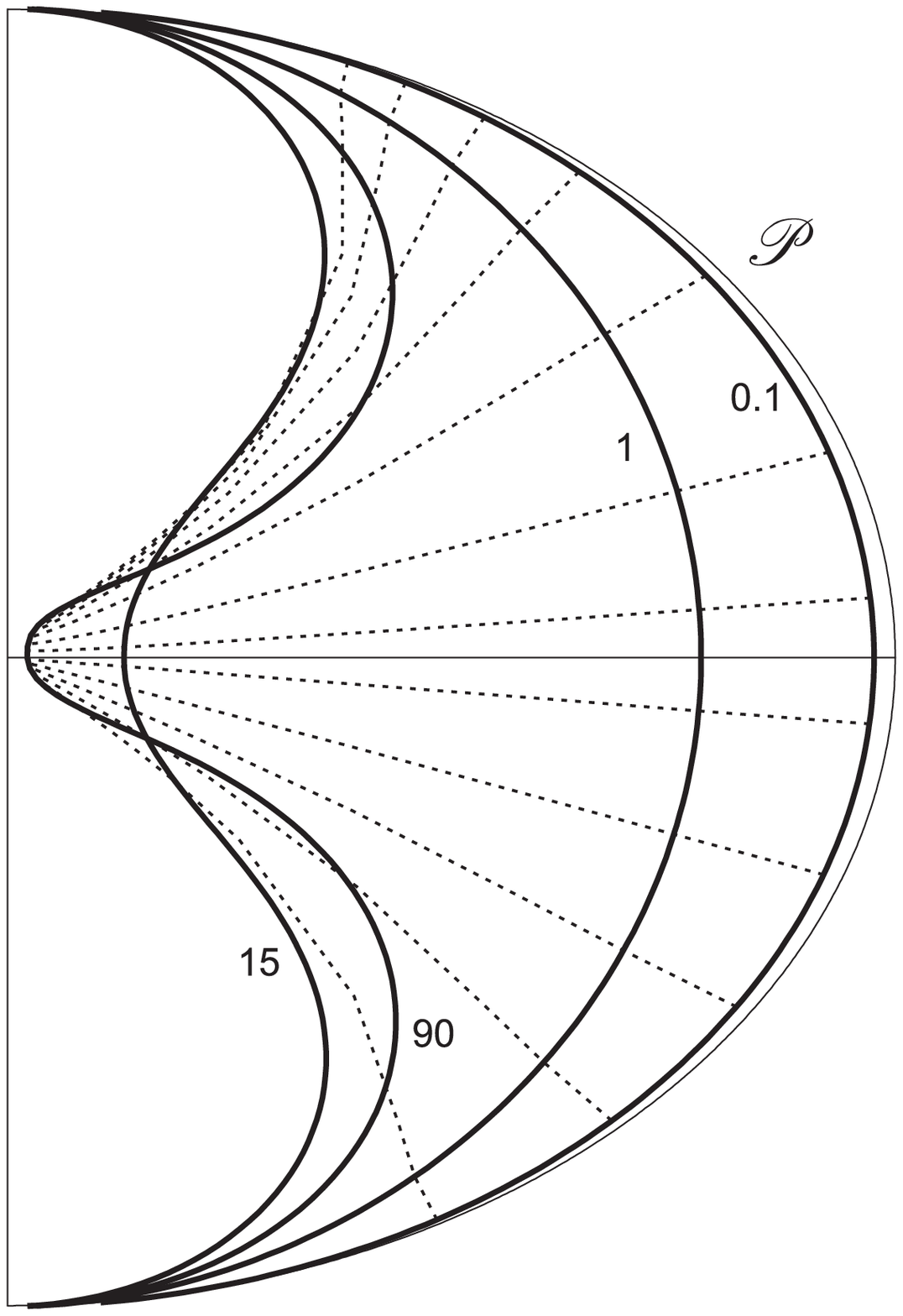}
\hfill
\end{center}
\caption{Azimuthal sections are plotted (thick solid curves) of
 the embedding surfaces in the Poincar\'e ball (denoted by the outer
 half-circle $\cal{}P$). Left panel: $Q=0$ (Schwarzschild black hole). Four cases
 of different ball curvature are shown; $\ell/M=0.1$, $1$, $2.4$, $4$.
 Right panel: $Q=M$ (maximally charged Reissner-Nordstr\"om black hole);
 $\ell/M=0.1$, $1$, $15$, $90$. The shape of the embedding curve
 resembles an arc if $\ell/M\rightarrow0$ (infinite 
 curvature), and it gets progressively more deformed when the curvature
 decreases. Dotted lines connect the points on the embedding
 curves with identical values of $r$. See the text for details.}
 \label{fig10}
\end{figure}

\begin{figure}[!tb]
\begin{center}
\hfill
\includegraphics[width=0.47\textwidth]{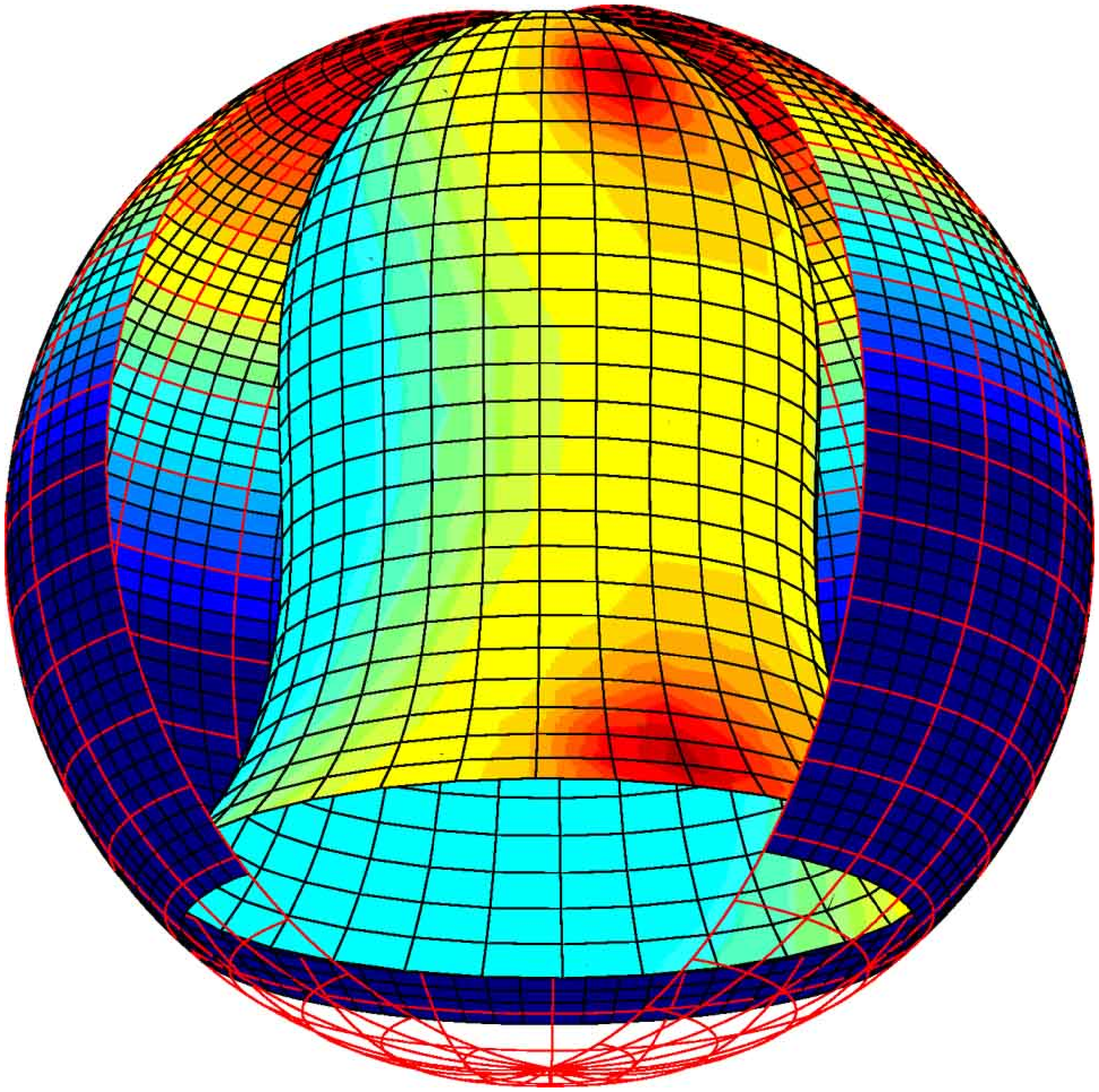}
\hfill
\includegraphics[width=0.47\textwidth]{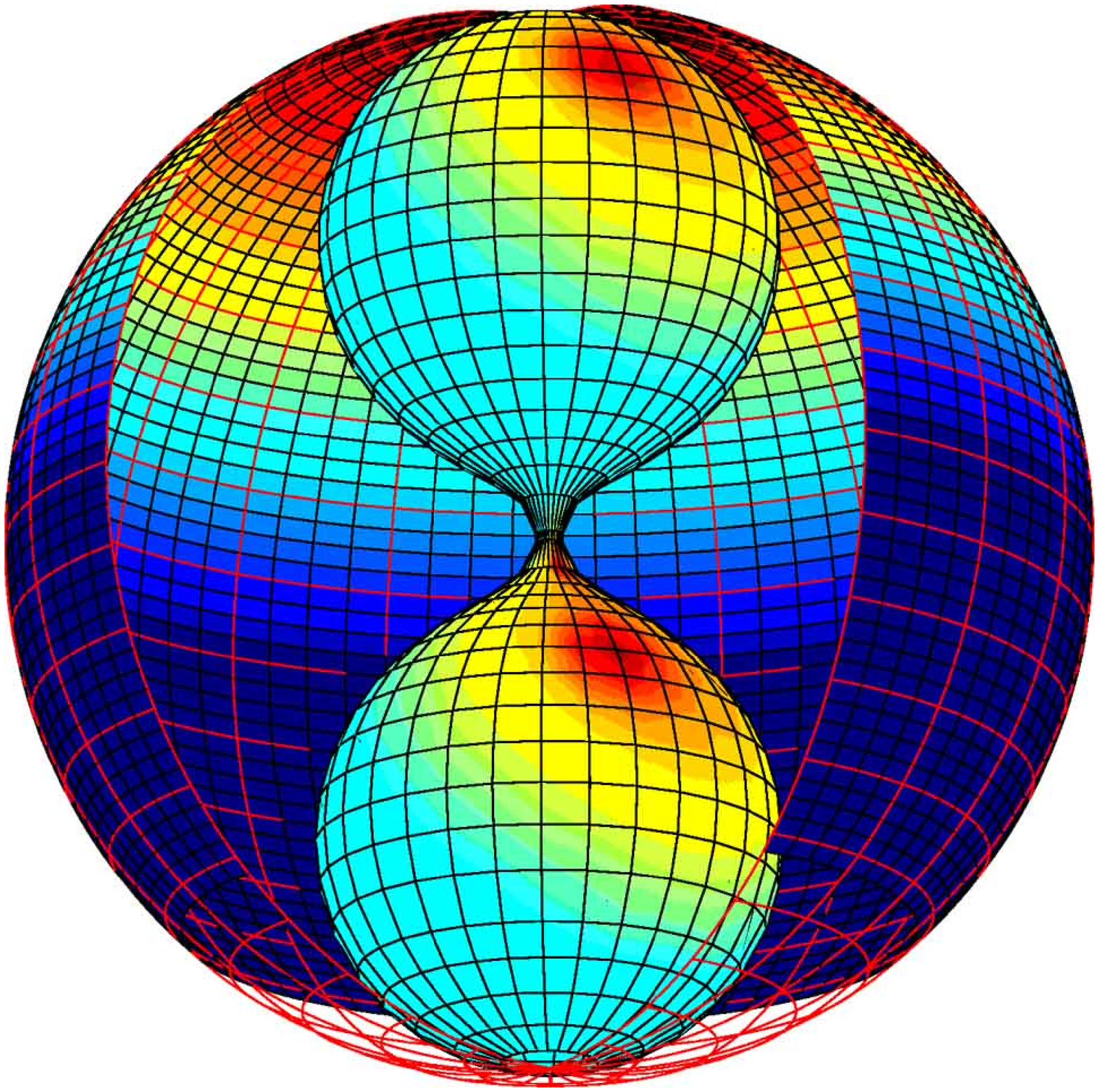}
\hfill
\end{center}
\caption{Poincar\'e ball embeddings.  Three-dimensional view of
 the embedding diagrams is obtained by rotating the azimuthal section
 around $z$-axis of symmetry.  Here we show two limiting cases
 corresponding to the graphs presented in fig.~\ref{fig10}.  Left panel:
 $Q/M=0$, $\ell/M=4$; right panel: $Q/M=1$, $\ell/M=90$.}
\label{fig20}
\end{figure}

We begin this section by discussing the behaviour of the embeddings in
two limiting cases, $r \rightarrow \infty$ and $r \rightarrow \rhor$.
Consider first the upper half space picture.  It is clear from the
analysis in section \ref{sec:embedding_equations} that $h \rightarrow
h_\infty$ as $r \rightarrow \infty$ where $h_\infty$ can be set to
arbitrary positive number.  Therefore the optical geometry defines an
asymptote to the Killing plane $h=h_\infty$ for large $r$.  This is
illustrated in fig.~\ref{fig:uh-schwarzschild} where a 2-dimensional
section of the Schwarzschild optical geometry is shown.  In the horizon
limit, $h\rightarrow 0$ as can be seen from eq.~\eqref{eq:h_expr2}
using the fact that $f(\xhor)=0$.  Thus the horizon is at infinity in
the hyperbolic space, and the embedding is defined all the way down to
(but not including) the horizon itself (see
fig.~\ref{fig:uh-schwarzschild} to illustrate this situation).

A particularly interesting aspect of the optical geometry of black
holes and relativistic stars is the appearance of circular light
orbits at constant radius. Such orbits correspond to necks and bulges
in the optical geometry (see e.g.\ \cite{abhp:antidesitter,karlovini}). 
The most well-known example of a neck is the one associated with
the circular orbit at $r=3M$ in the Schwarzschild geometry. The
existence of a neck implies that massless particles can be trapped in
the region inside the neck.
So, we turn to the manifestation of necks and bulges in the embedded
surface. The condition for a neck (or a bulge) occurring at some radius
$r_0$ is that the derivative $\tilde{r}'(r_0)=0$.  From
eq.~\eqref{eq:h_embed_cond} we see that this corresponds to $(q/h)'=0$,
which is precisely the condition that the surface is tangential to a ray
passing through the origin.  Clearly this is the same as being
tangential to the orbits of the Killing vector field $k_2$.  Since the
embedded surface has rotational symmetry, it is actually tangential to a
circular cone ruled by Killing orbits of the ray class in the upper half
space representation. The neck at $r=3M$ is shown in
fig.~\ref{fig:uh-schwarzschild} together with two associated Killing rays.
The appearence of necks can in fact be characterized in a 
simple way which is valid for any embedding with the rotational
symmetry, thus including both the flat and the hyperbolic 
embeddings.  Namely, the embedded surface has a neck
precisely if it touches a surface of revolution that is ruled by 
Killing orbits parallel to the symmetry axis.

Reinterpreting these results in the Poincar\'e ball picture we find that
the asymptotic optical geometry at infinity ($r \rightarrow \infty$) is
tangential to a Killing 2-sphere defined by north pole Killing orbits.
A 2-dimensional section of the Schwarzschild optical geometry embedded
in the ball is illustrated in fig.~\ref{fig:disk-schwarzschild}.  At the
horizon, the optical geometry again stretches out to infinity,
represented by the unit sphere with center at the origin, while at the
$r=3M$ neck the optical geometry touches a meridional Killing 
cylinder.\footnote{It may be worth to remark at this point that other
kinds of embeddings lead, in general, to a neck 
located away from the circular photon orbit
(cf.\ the case of usual $\Ebb^3$ embedding of the Schwarzschild
geometry); or the neck might disappear completely. It follows from 
the very definition of optical geometry that the 
neck coincides with the closed photon orbit \cite{abramowicz_etal:1988}.}

Thus the main conclusions in this section are as follows: (i)~asymptotic 
behaviour of the optical geometry for $r\rightarrow\infty$ is described by
Killing orbits of the north pole class; (ii)~necks and bulges occur when
the optical geometry is tangential to the meridional Killing cylinder;
and (iii)~the horizon is at infinity of hyperbolic space.

Finally, the actual shapes of Poincar\'e ball embeddings of the
Reissner-Nordstr\"om black hole are plotted in Figures
\ref{fig10}--\ref{fig20}.  In each graph we consider two values of the
charge for illustration: $Q=0$ (left panels) and $Q=M$ (right panels),
respectively. First, fig.~\ref{fig10} shows $\tilde\phi={\rm{const}}$
sections of the embedding surface.  The embedding shape touches the ball
in its northern pole corresponding to spatial infinity
($r\rightarrow\infty$), and at the other point corresponding to the
horizon ($r{\rightarrow}\rhor$).  Notice that the
latter point represents a circle on the embedding surface
($0\leq\phi<2\pi$) which gradually shrinks to the south pole of the ball
as $Q{\rightarrow}M$.  Thus, the embedding surface has no edge in the
limiting case of $Q=M$, as shown in the right panel.

Each panel contains four lines of the same embedding curve which refer
to different values of the curvature parameter $\ell/M$.  We recall that
the allowed range of $\ell/M$ is given by eq.~\eqref{eq:lm}: For $Q=0$, the
minimum radius of curvature is $\ell=4M$, corresponding to the embedding
curve crossing the ball perpendicularly at $r=2M$.  In the strong
curvature limit ($\ell/M \rightarrow0$), the embedding surface
asymptotically becomes spherical and hence approaches the ball.  The
situation is slightly more complicated for $Q=M$ and large curvature,
when the neck develops on the embedding surface.
A three-dimensional view of the embedding is obtained by rotating the
embedding curve around symmetry axis; two such examples are plotted in
fig.~\ref{fig20}.

Notice that the right hand panels of figures \ref{fig10} and
\ref{fig20}, depicting the extremal case, are symmetric under reflection
in a horizontal plane through the neck. This is not an accident: It is
known \cite{couch:1984} that the extremal Reissner-Nordstr\"om black
hole admits a discrete conformal isometry that interchanges the region
$M<r<2M$ with the region $r>2M$. For the optical geometry this
transformation becomes a true isometry. Using the tortoise coordinate
$r_{\star}$ it is simply a reflection $r_{\star} \rightarrow -r_{\star}$ 
through the neck, which is situated at $r_{\star}=0$.

\section{Conclusions and future work}
We have argued in this paper that the Poincar\'e ball embedding is a
natural and useful tool to study global topological properties of
spaces with negative curvature, for example the optical space of a
black hole geometry.  One particular subject that we plan to study
using the Poincar\'e ball embeddings is the problem of topology of the
event horizon in the optical space corresponding to the
Reissner-Nordstr\"om solution.  In the optical space, the event horizon
is always located at infinity, corresponding to the surface of the
Poincar\'e ball in the embedding.  It is interesting to note that while
for non-extremal ($Q<M$) black holes the region occupied by the horizon
is finite (see fig.~\ref{fig20}, left panel), in the case of an
extremal black hole it shrinks to a single point (fig.~\ref{fig20},
right panel).   This is because the optical geometry of the extremal
black hole is asymptotically flat at both ends. 

\ack
VK acknowledges hospitality of SISSA (Trieste), and support 
from GACR\,205/00/1685 and 202/02/0735.

\end{document}